\documentclass[conference]{IEEEtran}
\IEEEoverridecommandlockouts
\usepackage{acronym}
\usepackage{amsmath,amssymb,amsfonts}
\usepackage{algorithmic}
\usepackage{booktabs}
\usepackage{cite}
\usepackage{comment}
\usepackage{graphicx}
\usepackage{multirow}
\usepackage{tabularx}
\usepackage{textcomp}
\usepackage{xcolor}
\usepackage{xspace}

\usepackage{hyperref}
\hypersetup{
    colorlinks=true,
    citecolor=green,
    filecolor=black,
    linkcolor=red,
    urlcolor=blue
}
\usepackage{cleveref}

\def\BibTeX{{\rm B\kern-.05em{\sc i\kern-.025em b}\kern-.08em
    T\kern-.1667em\lower.7ex\hbox{E}\kern-.125emX}}

\acrodef{acpf}[ACPF]{alternating current power flow}
\acrodef{acopf}[ACOPF]{alternating current optimal power flow}
\acrodef{ai}[AI]{artificial intelligence}
\acrodef{dcopf}[DCOPF]{direct current optimal power flow}
\acrodef{scacopf}[SC ACOPF]{security-constrained alternating current optimal power flow}
\acrodef{cpu}[CPU]{central processing unit}
\acrodefplural{cpu}[CPUs]{central processing units}
\acrodef{gpu}[GPU]{graphical processing unit}
\acrodefplural{gpu}[GPUs]{graphical processing units}
\acrodef{gui}[GUI]{graphical user interface}
\acrodef{hpc}[HPC]{high-performance computing}
\acrodef{llm}[LLM]{large language model}
\acrodefplural{llm}[LLMs]{large language models}
\acrodef{pdf}[PDF]{portable document format}
\acrodef{rag}[RAG]{retreival augmented generation}
\acrodef{simd}[SIMD]{single-instruction multiple-data}
\acrodef{uc}[UC]{unit commitment}
\acrodefplural{uc}[UCs]{unit commitments}
\acrodef{dduc}[DDUC]{data-driven unit commitment }
\acrodef{sc}[S.C.]{South Carolina}

\newcommand{\exago}{ExaGO\xspace}
\newcommand{\pflow}{PFLOW\xspace}
\newcommand{\opflow}{OPFLOW\xspace}
\newcommand{\scopflow}{SCOPFLOW\xspace}
\newcommand{\sopflow}{SOPFLOW\xspace}

\begin{document}

\title{Integrating Agentic Artificial Intelligence with High-Performance Computing for Grid Planning 
\thanks{This manuscript has been authored in part by UT-Battelle, LLC, under contract DE-AC05-00OR22725 with the US Department of Energy (DOE). The US government retains and the publisher, by accepting the work for publication, acknowledges that the US government retains a non-exclusive, paid-up, irrevocable, world-wide license to publish or reproduce the submitted manuscript version of this work, or allow others to do so, for US government purposes. DOE will provide public access to these results of federally sponsored research in accordance with the DOE Public Access Plan (https://energy.gov/doe-public-access-plan).}
}

\author{\IEEEauthorblockN{Samim Konjicija}
\IEEEauthorblockA{\textit{Faculty of Electrical Engineering} \\
\textit{University of Sarajevo}\\
Sarajevo, Bosnia and Herzegovina \\
skonjicija@etf.unsa.ba}
\and
\IEEEauthorblockN{Slaven Pele\v{s}}
\IEEEauthorblockA{\textit{Computational Science and Engineering Division}}
\textit{Oak Ridge National Laboratory}\\
Oak Ridge, TN, USA \\
peless@ornl.gov}

\maketitle

\begin{abstract}
We present AgentiGrid, an agentic \ac{ai} framework that integrates \ac{llm} intelligence and \ac{hpc} to streamline and accelerate the multi-scenario power flow studies. AgentiGrid is an autonomous decision-making agent that proposes parameter modifications, invokes
analyses through \ac{hpc} analysis toolkit \exago, interprets results, and determines subsequent actions. \exago provides multiple power flow applications that can perform deterministic, stochastic and security constrained optimal power flow analyses. AgentiGrid provides backends to multiple \acp{llm} (OpenAI, Anthropic, Ollama, and Ollama cloud) augmented with context specific and task specific prompts. Key features include interactive mid-search steering, goal-type-aware post-search analysis, and concurrent variant exploration for power flow optimization. A Streamlit-based graphical launcher provides real-time visualization of iteration progress and generates reports in natural language. AgentiGrid is capable of autonomously converging transmission constrained \ac{acopf} in under 20 iterations, with near-perfect reliability.
\end{abstract}

\begin{IEEEkeywords}
Agentic AI, generative AI, power systems, optimal power flow, high-performance computing
\end{IEEEkeywords}

\section{Introduction}

Power system engineers regularly face analysis tasks that are hard to formulate as mathematical optimization problems but can be easily expressed in naturally language: ``find the load level at which the network becomes infeasible,'' ``identify the weakest transmission corridor under stress,'' or ``find a generation dispatch that balances cost and voltage quality'' \cite{Dong2024}. A human expert approaches these tasks through iterative exploration -- running a simulation, examining results, forming hypotheses, adjusting parameters, and repeating. This expert workflow is often time-consuming. The emergence of parallel computing for power systems has accelerated numerical analyses \cite{Khaitan2013} thus leaving human decision making processes as the main bottleneck.

Recent advances in \acp{llm} have demonstrated strong capabilities in structured reasoning, code generation, and multi-step problem solving. The broader potential of \acp{llm} for power system planning — spanning fault diagnosis, load forecasting, and optimization — has recently been surveyed, revealing both promising capabilities and open challenges related to domain-specific reliability \cite{Dong2024}. As of now, there exist at least 5 different agentic AI solutions specifically targeting power systems operations and planning. Some of them \cite{cheng2025gaia} function only as an advisor, analyzing the model grids and providing suggestions to the human operator. Some, like \cite{zhang2025poweragent} and \cite{jin_gridmind_2025}, are capable of running simulations in the background and, for instance, identifying which N-1 contingencies or which load increment volumes are harmful. More advanced agentic systems, such as \cite{zhang2025grid}, can resolve a harmful contingency and contain continuous improvement loops. The agents that are capable of autonomously running power flow simulations usually follow the legacy sequential process. The agent orchestrator receives a scenario and sends to an existing power flow package, e.g. PandaPower or PowerWorld.

Recent surveys confirm that existing \ac{llm}-based power system agents predominantly follow a sequential, single-simulation-per-step paradigm, leaving parallel execution largely unexplored \cite{Sarwar2025}. This sequential design incurs high token and time costs to resolve a contingency or process a scenario. Since agentic AI is meant to relieve humans of iterative, slow tasks, embedding parallelism into the workflow is a logical next step \cite{Gonzalez2021}. AgentiGrid extends this line of work by incorporating \exago \cite{peles2026exago}, a high-performance power grid optimization library developed at Oak Ridge National Laboratory for parallel execution on clusters and cloud systems, wrapping its simulation applications in an \ac{llm}-driven search loop that analyzes multiple scenarios or contingencies in parallel and postprocesses the resulting outputs.

The main contributions of this paper are:
\begin{itemize}
    \item A three-layer architecture separating the \ac{llm} agent loop, simulation orchestration, and \exago, supporting six power flow and optimal power flow applications through a unified interface.
    \item An application-aware command system with deterministic validation, runtime feedback to the \ac{llm}, and application-specific command vocabularies covering transformer tap ratios, bus shunt susceptances, and phase shift angles.
    \item A prompt architecture encoding search history, network metadata, goal context,
          and operator directives, with per-application guidance sections.
    \item Interactive steering allowing operators to inject directives into a running
          search and pause/resume at iteration boundaries without restarting.
    \item Concurrent variant exploration for power flow-based optimization, amortizing \ac{llm}
          round-trip latency across multiple parallel simulations.
    \item A Streamlit-based graphical launcher with real-time monitoring, multi-objective
          visualization, and natural language report generation.
\end{itemize}

An overview of the three-layer architecture is given in \cref{sec:architecture}. 
We discuss use of an \ac{llm} as an optimization engine in \cref{sec:optimizer}. In \cref{sec:results} we present results of our experiments, and we summarize our findings and outline next steps in \cref{sec:conclusion}. 

\section{System Architecture Overview}
\label{sec:architecture}

\begin{figure}[htbp]
    \centering
    \includegraphics[width=\columnwidth]{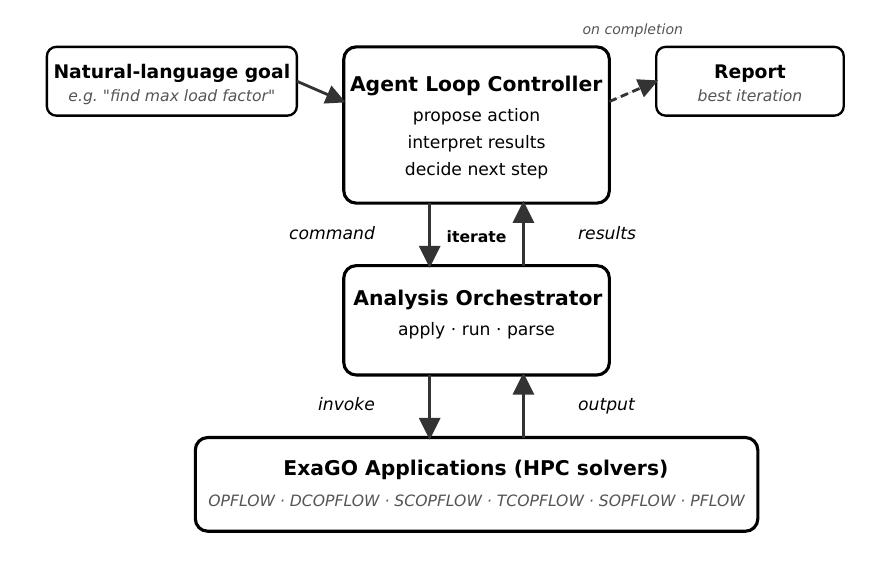}
    \caption{AgentiGrid system architecture and its data flow per iteration.}
    \label{fig:architecture}
\end{figure}

AgentiGrid architecture is organized into three layers:
\begin{itemize}
    \item \textbf{Agent Loop Controller} manages the iterative cycle, assembles
          \ac{llm} prompts from the search goal, journal history, latest results and active
          steering directives, parses \ac{llm} responses into structured actions, and records
          results in the search journal.
    \item \textbf{Analysis Orchestrator} validates and applies modification
          commands to working copies of the base case network file, invokes \exago binaries
          as subprocesses, passes output to application-specific parsers, and maintains the search journal.
    \item \textbf{\exago Applications} are \ac{hpc} simulation engines for power flow, optimal power flow, security constrained optimal power flow, and stochastic optimal power flow analyses. They are invoked as external executables.
\end{itemize}
The three layers communicate through well-defined protocols and do not depend on each other implementations.
This architecture and its workflow is schematically described in \cref{fig:architecture}.

One complete search iteration proceeds as follows: The Agent Loop Controller receives the power system case information in Matpower format and a request for specific action in natural language through the application interface. The controller then  assembles the prompt and passes it to \ac{llm}. Next, the \ac{llm} generates a command in JSON format and passes it to the Analysis Orchestrator. The orchestrator validates the proposed instruction against the power system case data, creates a working copy of that data in a separate timestamped directory, and applies validated commands to that copy. The orchestrator then invokes the appropriate \exago application to perform numerical analysis. Upon analysis completion, the orchestrator extracts the structured data from \exago's output, checks constraint violations against the actual limits encoded in the modified case file, updates a search journal, and sends a compact summary for the Agent Loop Controller to incorporate in the next prompt. The controller assembles the next prompt and starts another iteration. This loop repeats until the \ac{llm} declares successful completion, a critical error occurs, or the maximum iteration count is reached.

After the search loop is exited, the final \ac{llm} call classifies the search goal type and identifies the best iteration, providing a goal-aware framing for the results report.

\subsection{Agent Loop Controller}
\label{sec:controller}

The \ac{ai} agent governs the grid planning analyses through the Agent Loop Controller, which is a standalone Python module. This module generates prompts for the \ac{llm} agent based on user input and prior iterations, sends commands to the Analysis Orchestrator, and manages the post-search analysis. The main components of this module are user fronted, \ac{llm} backend, prompt generator, interactive steering, and report generator.

\subsubsection{User Frontend}

AgentiGrid provides command line and a web-based \ac{gui}. Both interfaces allow user to select power grid case file, analysis type, \ac{llm}, \exago application and additional configuration options. User enters requests in plain text format. Several predefined (modifiable) prompts are available as the starting point.

\subsubsection{LLM Backend}

All \ac{llm} providers are accessed through a common abstract interface exposing \texttt{complete()}, \texttt{name()}, and \texttt{supports\_json\_mode()} methods. Concrete implementations exist for OpenAI, Anthropic, Ollama, and Ollama cloud. \ac{llm}'s temperature is kept low (0.2--0.4) to encourage consistent structured output.

\subsubsection{Prompt Architecture}
\label{sec:prompt}

At each iteration the Agent Loop Controller assembles an \ac{llm} prompt from up to seven sections. Some of the sections are set at the beginning and stay the same during the analysis (static), some are appended as new data is generated (grow), some are periodically replaced when new data is generated (refreshed), and some are created upon certain events (conditional). These are:

\begin{itemize}
    \item \textbf{System Prompt (static):}
    Role definition, full command schema with parameter types and validation rules,
    power grid model, and response format.
    \item \textbf{Network Metadata (static)} includes power system configuration options such as
    slack bus designations, must-run generators, and offline generators. This section is created once at session start.
    \item \textbf{Goal Statement (static)}
    is  user's natural language goal, preserved verbatim throughout the search.
    \item \textbf{Search Journal (grows)}
    is a compact table with one row per iteration with following columns: iteration number, modification
    description, objective value, feasibility, voltage range, and maximum line loading.
    \item \textbf{Latest Results (refreshed)}
    are a compact summary from the most recent simulation (15-30 lines of plain text).
    \item \textbf{Operator Directives (conditional)}
    could be injected by user via augment or replace mode (see \cref{sec:steering}).
    \item \textbf{Error/Warning Feedback (conditional)}
    is generated by the Analysis Orchestartor and sent to Agent Loop Controller. The feedback may include command validation errors, JSON parse failures, and application-specific warnings
    from the prior iteration.
\end{itemize}
The prompt composed in this way is passed at each agent iteration.

\subsubsection{Interactive Steering}
\label{sec:steering}

Users can inject directives into a running search without stopping it. The mechanism uses a thread-safe queue to inject command to modify the case study. User can add a steering command through command-line or \ac{gui}. 
Two directive modes are supported -- augment and replace. In augment mode, the directive is added alongside the current goal. In replace mode, the new directive clears all prior directives and becomes the sole active goal. Active directives are injected as a dedicated Operator Directives section of the \ac{llm} prompt (see \cref{sec:prompt}).

\subsubsection{Report Generator}

Reports are generated in \ac{pdf} using ReportLab Python module with full Unicode support.
Report sections include: title page, executive summary with goal-type-aware framing,
convergence and voltage range charts (exported from Plotly via kaleido), base-case versus
best-solution comparison table, per-iteration log, and steering history. Charts are
rendered identically in the \ac{gui} and the \ac{pdf} report by sharing the same Plotly figure-builder
functions.

\subsection{Analysis Orchestrator}
\label{sec:orchestrator}

The Analysis Orchestrator is an intermediary between the \ac{llm} agent and the \ac{hpc} application. It consists of the following components:

\subsubsection{Modification Engine}

The Modification Engine is a deterministic Python module that receives JSON commands from the Agent Loop Controller and applies them to power flow case files in Matpower format, drawing on the supported command vocabulary listed in Table~\ref{tab:commands}. Before applying a command, the engine validates it against the actual network topology; invalid commands are rejected, and the resulting error message is fed back to the \ac{llm} in the next prompt.

\begin{table}[htbp]
\caption{Supported Modification Commands}
\label{tab:commands}
\begin{center}
\begin{tabularx}{\columnwidth}{lX}
\toprule
\textbf{Command} & \textbf{Description} \\
\midrule
\texttt{set\_load}              & Set active/reactive load at a bus \\
\texttt{scale\_load}            & Scale loads in an area, zone, or bus \\
\texttt{scale\_all\_loads}      & Scale all system loads uniformly \\
\texttt{set\_gen\_status}       & Enable or disable a generator \\
\texttt{set\_gen\_dispatch}     & Set generator active power output \\
\texttt{set\_gen\_voltage}      & Set generator voltage setpoint \\
\texttt{set\_branch\_status}    & Enable or disable a branch \\
\texttt{set\_branch\_rate}      & Modify thermal rating \\
\texttt{set\_cost\_coeffs}      & Modify generator cost curve \\
\texttt{set\_bus\_vlimits}      & Set voltage bounds on a single bus \\
\texttt{set\_all\_bus\_vlimits} & Set voltage bounds on all buses \\
\texttt{set\_tap\_ratio}        & Set transformer tap ratio \\
\texttt{set\_shunt\_susceptance}& Set bus shunt susceptance \\
\texttt{set\_phase\_shift\_angle}& Set phase-shifting transformer angle \\
\texttt{scale\_load\_profile}   & Scale per-period CSV load profiles (TCOPFLOW) \\
\texttt{scale\_wind\_scenario}  & Scale wind scenario CSV (SOPFLOW) \\
\bottomrule
\end{tabularx}
\end{center}
\end{table}

\subsubsection{Simulation Executor}

\exago applications are invoked through their command line interfaces. A configurable timeout guard prevents
hanging simulations. The executor also provides a parallel execution method using
Python's \texttt{ThreadPoolExecutor} for concurrent variant execution in power flow explore
mode (see \cref{sec:optimizer}).

\subsubsection{Results Parser}

For each \exago application, there is a dedicated parser extracting the objective value, convergence
status, bus voltages, generator dispatch, branch flows, and constraint violations.
Violation checking uses the actual $V_{\min}$, $V_{\max}$ limits from the modified
network rather than the baseline case, ensuring violations are reported accurately in case that
the \ac{llm} changes limits dynamically.
The parser layer produces two outputs per simulation: a full structured Python data structure
for visualization queries, and a compact 15--30 line text summary for \ac{llm} prompt
inclusion.

\subsubsection{Search Journal}

The search journal is an in-memory data structure with one entry per iteration, recording:
iteration number, modification description, commands applied, objective value, feasibility,
violation count, voltage range, maximum line loading, total generation and load, \ac{llm}
reasoning, modification mode, and any active steering directive. Multi-objective tracking
adds a \texttt{tracked\_metrics} dictionary of per-iteration values for each registered
objective.

\subsubsection{Goal Classification and Post-Search Analysis}
\label{sec:goal-classification}

After the search loop, a final \ac{llm} call classifies the search goal type and identifies the
best iteration. Four goal types are recognized:
\begin{itemize}
    \item \textbf{cost\_minimization}: best iteration has lowest feasible cost.
    \item \textbf{feasibility\_boundary}: best iteration is the feasible configuration
    closest to the infeasibility boundary.
    \item \textbf{constraint\_satisfaction}: best iteration best satisfies specified
    constraints.
    \item \textbf{parameter\_exploration}: best iteration is the most informative feasible
    configuration.
\end{itemize}

Goal-type-aware framing is essential for correct reporting. For example, in a feasibility boundary
search, generation cost increases as load grows, so the na{\"i}ve ``best = lowest cost''
heuristic would incorrectly identify the base case as best. If classification fails, the
system falls back to cost minimization.

\subsection{ExaGO Applications}
\label{sec:exago}

All physics based analyses are performed using applications from \exago package. The list of \exago applications is given in \cref{tab:exago-apps}.

\begin{table}[htbp]
\caption{\exago applications.}
\label{tab:exago-apps}
\begin{center}
\begin{tabularx}{\columnwidth}{lX}
\toprule
\textbf{Application} & \textbf{Description} \\
\midrule
\pflow    & alternating current power flow simulation \\
\opflow   & alternating current optimal power flow analysis \\
DCOPFLOW & direct current optimal power flow analysis \\
TCOPFLOW & multiperiod optimal power flow analysis \\
SCOPFLOW & security constrained optimal power flow analysis \\
SOPFLOW  & stochastic optimal power flow analysis (optionally can include security constraints) \\
\bottomrule
\end{tabularx}
\end{center}
\end{table}

\section{Using LLM as the Optimizer}
\label{sec:optimizer}

Power flow analysis implemented in \pflow module is a forward steady-state simulation. The Newton-Raphson solver finds the power
flow solution for a given network state. AgentiGrid can use \pflow in a sampling-based optimization mode
where the \ac{llm} serves as the optimizer, proposing dispatch changes, evaluating feasibility,
and deciding next steps.

\subsection{Explore/Select Actions}
\label{sec:explore}

Traditional sequential \pflow search evaluates a single configuration per iteration.
Each iteration incurs one \ac{llm} inference call ($\sim$20--30\,s), while the
corresponding simulation requires only $\sim$0.02\,s. The \ac{llm} call therefore
dominates total runtime. Concurrent \pflow addresses this bottleneck by introducing
two new actions, explore and select. In the explore action, the \ac{llm} generates
between 2 and 8 independent variant command sets within a single iteration; these
variants are simulated concurrently using Python \texttt{ThreadPoolExecutor}, and the
resulting outcomes are summarized via Pareto front analysis. In the select action, the
\ac{llm} designates one variant as the new current network state, and this decision,
along with the full variant comparison metadata, is recorded in the journal. Each
explore-select cycle is counted as a single iteration toward the overall iteration
budget, yet evaluates multiple configurations, thereby amortizing the \ac{llm} latency
across those simulations.

\subsection{Pareto Front Computation}

The Pareto filtering procedure applies standard Pareto dominance: feasible candidates
dominate infeasible ones; among feasible candidates, $A$ dominates $B$ if $A$ is at
least as good as $B$ on all objectives and strictly better on at least one. Objective
directions (minimization or maximization) are configurable. By default, variants are
ranked by generation cost, which is minimized.
Prior to execution, variants consisting entirely of invalid commands (e.g., dispatching
the slack bus) are identified and excluded from simulation. No subprocess is launched
for such a variant, and it is instead marked as rejected in the results table.

\section{Experimental Results}
\label{sec:results}

We evaluate AgentiGrid on ACTIVSg200~\cite{TAMUgrid}, a synthetic 200-bus network representing a regional transmission system in Illinois. The network comprises 246 branches (including 197 transformers), 200 loads, and 49 generators (40 in service), with nominal voltages ranging from 13.8\,kV to 230\,kV. Its low node degree (1.23 edges per node) and frequent transmission-to-distribution voltage transitions make it a non-trivial testbed for agent reasoning. All the analyses were run on a workstation with Intel Core i7-13650HX with 14 cores, and with 32 GB of RAM.

\subsection{Experimental Design}
\label{sec:experimental-design}

We designed three goal strings, each using different \exago executable to perform different optimal power flow analyses:

\begin{itemize}
  \item \textbf{\pflow} (sampling-based optimization): ``Reduce the total generation cost by adjusting
        generator dispatch while maintaining feasibility.'' The base case is already
        economically dispatched (\$27{,}564), so this probes whether the \ac{llm}-as-optimizer
        can improve a near-optimal point.
  \item \textbf{SCOPFLOW} (security-constrained analysis): ``Systematically test N-1 single-line outages to
        identify the most critical lines, ranked by cost increase, voltage violation, or
        infeasibility.''
  \item \textbf{SOPFLOW} (stochastic analysis): ``Using the wind scenario file, evaluate whether
        the grid can handle all wind generation scenarios simultaneously, and identify the buses
        most affected by wind variability.''
\end{itemize}


The four backends---Anthropic Claude Sonnet 4.6, OpenAI GPT-5.4, Zhipu GLM-5.1, and DeepSeek-V4-Pro (the latter two via Ollama Cloud)---received an identical goal
string per application, including a standing instruction to keep exploring modified problems rather
than stop when the solver reports a converged but uninformative result. Holding the
prompt constant across backends ensures that any observed differences in behavior
are due to the models rather than to variation in prompting. All runs had a
maximum iteration budget of 20 and \ac{llm} temperature was set to 0.3. SCOPFLOW and SOPFLOW
apps used \exago's parallel EMPAR backend. \pflow used the Newton--Raphson
solver to compute steady-state power flow solution.

\subsection{Key Findings}

Table~\ref{tab:xresults} reports the twelve searches. All four backends completed all three tasks without fatal JSON parsing failures. The backends separate on (i) \emph{search strategy}, i.e.~how efficiently each converts its iteration budget into useful simulations, and (ii) on the \emph{depth} of the conclusions reached.

\begin{table*}[!htb]
\caption{Cross-Application Results on ACTIVSg200 (Identical Goal Prompts per Application)}
\label{tab:xresults}
\begin{center}
\begin{tabular}{l l r r r l}
\toprule
\textbf{Application} & \textbf{Backend} & \textbf{Iter.} & \textbf{t (s)} & \textbf{Tokens} & \textbf{Principal result} \\
\midrule
\multirow{4}{*}{\pflow}
 & Claude Sonnet 4.6 &  8 & 130 &  70,793 & No gain on base (\$27,564) \\
 & GPT-5.4           & 18 & 159 & 164,730 & No gain on base (\$27,564) \\
 & GLM-5.1           & 16 & 580 & 176,039 & Performed unit commitment instead, 36.9\% cost reduction (\$17,398) \\
 & DeepSeek-V4-Pro   & 15 & 342 & 197,354 & Performed unit commitment instead, 34.9\% cost reduction (\$17,941) \\
\midrule
\multirow{4}{*}{SCOPFLOW}
 & Claude Sonnet 4.6 &  7 & 205 & 164,559 & 187$\to$189 critical (+21.6\%); no infeasible case found \\
 & GPT-5.4           & 18 & 112 & 150,962 & 187$\to$189 critical (+21.6\%); 29$\to$30 infeasible \\
 & GLM-5.1           &  7 & 188 &  72,705 & 187$\to$189 critical (+21.6\%); 29$\to$30 infeasible \\
 & DeepSeek-V4-Pro   & 11 & 292 & 177,899 & 187$\to$189 critical (+21.6\%); 29$\to$30 infeasible \\
\midrule
\multirow{4}{*}{SOPFLOW}
 & Claude Sonnet 4.6 &  3 &  29 &  21,408 & All solutions feasible; bus 152 most affected\\
 & GPT-5.4           &  3 &  12 &  18,585 & All solutions feasible; identified 15 most sensitive buses, bus 152 most affected\\
 & GLM-5.1           &  3 &  19 &  19,545 & All solutions feasible; identified 15 most sensitive buses, bus 152 most affected\\
 & DeepSeek-V4-Pro   &  3 &  20 &  19,183 & All solutions feasible; bus 152 most affected\\
\bottomrule
\end{tabular}
\end{center}
\end{table*}

\textbf{\pflow: a near-null task (in)correctly recognized.} We asked the agents a trick question: optimize a case that was already economically dispatched. Sonnet 4.6 and GPT 5.4 correctly recognized optimal solution and declared a failure because they did not reduce the cost. Sonnet reached that conclusion with fewer iterations, which projected to lower cost (fewer tokens) and slightly shorter time compared to GPT. GLM 5.1 and DeepSeek 4 ignored instruction to ``adjust generator dispatch'' and performed unit commitment instead. Since the test case had large spinning reserve, they obtained 36.9\% and 34.9\% cost reductions, respectively. DeepSeek got to the solution significantly faster but at slightly higher cost in terms of tokens.

\textbf{SCOPFLOW: contingency ranking is robust; infeasibility detection is not.} Every backend correctly identified line 187$\to$189 connecting the nuclear power plant as the most economically critical N-1 outage, with all four agreeing on the +21.6\% cost increase. Where the models diverged was in detecting the \emph{infeasible} contingencies that could possibly lead to blackouts. Sonnet did not identify 29$\to$30 line outage as critical and all four \acp{llm} failed to uncover a second infeasible outage (15$\to$16) probably because they did not screen deep enough into the line loading ranked list. 

\textbf{SOPFLOW: Parallel computing helps improve agent efficiency.} By using parallel computing we can make evaluation of all stochastic scenarios a single evaluation step by the \ac{llm} agent, thus reducing the number of overall iterations to solution and the number of tokens. All agents completed the stochastic analysis task in only 3 steps, finding feasible solution for each generation scenario. All agents identified bus 152 as the most sensitive to voltage violations. GPT 5.4 and GLM 5.1 have found 15 buses with similar sensitivities, which is what grid planners would expect, while Sonnet 4.6 and DeepSeek 4 only reported the most sensitive one.

\subsection{Results Summary}

Overall, the preliminary results were encouraging. All four \ac{llm} agents performed
the three tasks well. The only significant shortcoming was the failure of Sonnet 4.6
to identify any contingencies that lead to infeasible optimal power flow solutions.
Other, less critical issues that we observed can be addressed with a more comprehensive
\ac{rag} knowledge base. Adding a dictionary with more precise interpretations of
economic dispatch and unit commitment would likely lead GLM-5.1 and DeepSeek-V4-Pro to
reach correct answers when using sampling-based optimization with \pflow. Similarly, every
power system engineer would recognize that an infeasible optimal power flow case often
corresponds to a situation leading to a potential blackout, which is a more serious
situation than a cost increase due to redispatch. Such engineering experience needs to
be incorporated into \ac{rag} so that an \ac{llm} agent can prioritize analyses
regardless of the order in which requirements are listed in the goal statement. In the
security-constrained analysis with the SCOPFLOW app, agents prioritized cost over
feasibility simply because cost was listed before feasibility in the prompt (see
\cref{sec:experimental-design}). Also, \acp{llm} struggled to determine how to search
for the ``most critical lines,'' causing them to miss the line 15$\to$16 contingency,
which leads to an infeasible solution. Including more specific criteria for line
contingency rankings in \ac{rag} could address this issue.

Another possible improvement is to leverage \ac{hpc} more extensively when available. A
broader contingency sweep in the security-constrained analysis with \scopflow would
uncover the infeasible optimal power flow solution associated with the line 15$\to$16
failure. Parallel computing proved effective in the stochastic analysis with \sopflow,
where all \ac{llm} agents reached the correct solution in only three iterations because
all the data for the analysis was generated in a single \ac{llm} agent iteration.
In the \sopflow case, additional criteria in \ac{rag} on how to identify
``most affected buses'' would be beneficial. In our trials, Sonnet 4.6 and
DeepSeek 4 interpreted that instruction as asking them to find a single most
affected bus. In engineering practice, however, all buses exhibiting similarly severe
overvoltages would be considered equally important.

\begin{figure}[htbp]
\centering
\includegraphics[width=\columnwidth]{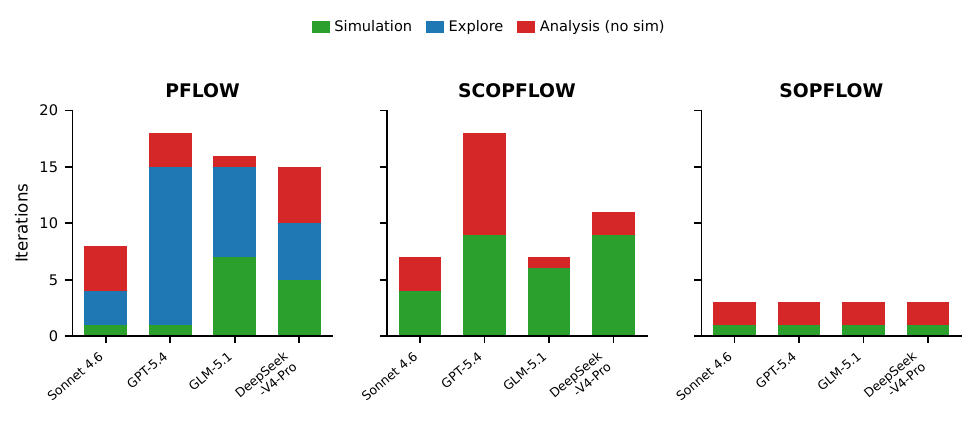}
\caption{Decomposition of each search's iteration budget into productive
simulation, concurrent explorations, and analysis-only steps. Analysis and exploration-heavy behavior, most pronounced for GPT-5.4, reflects repeated requests for network detail not
present in the compact result summary.}
\label{fig:iterations}
\end{figure}

Our preliminary results show that GPT-5.4 takes the largest number of steps to reach
the solution across all three problems, but it also provides the most accurate
solutions. In contrast, Claude Sonnet 4.6 takes the fewest iterations of all four
agents but fails to identify a critical contingency. A summary of the number of
iterations to solution for each \ac{llm} agent is given in \cref{fig:iterations}.

\begin{figure}[htbp]
\centering
\includegraphics[width=\columnwidth]{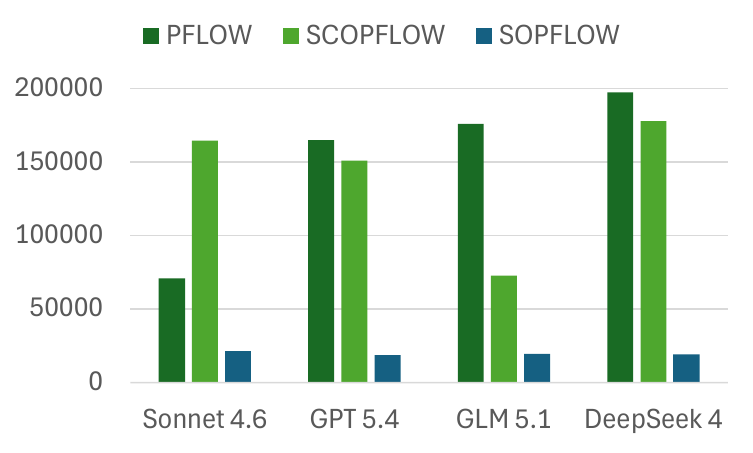}
\caption{Total token consumption per search, by backend and application.}
\label{fig:tokens}
\end{figure}

We found that the cost of the analysis, measured in tokens, varies significantly from
problem to problem, from agent to agent, and even from iteration to iteration. A
summary of the number of tokens used by each agent over the three problems is given in
\cref{fig:tokens}.

\section{Conclusion}
\label{sec:conclusion}

AgentiGrid provides a practical framework for \ac{llm}-driven iterative power grid
simulation, bridging natural language problem specification and production-grade
\exago solvers. The architecture supports six \exago applications through a unified
interface with application-aware command validation, prompt construction, and results
parsing.

Currently, AgentiGrid is intended as a research platform to explore opportunities for
using agentic \ac{ai} in combination with parallel computing for power systems
planning. Our current implementation supports two types of parallelism: launching
concurrent \ac{llm} explorations, as in the \pflow analysis (\cref{sec:optimizer}), and using \ac{hpc}
executables for physics-based analyses. Our preliminary experiments suggest this is a
promising direction to explore. Furthermore, we find that the ``\ac{llm} as optimizer''
paradigm is practically viable for power system analysis tasks that are difficult to
express as formal optimization problems.

All of our preliminary experiments were done for a 200-bus optimal power flow case.
The next step is to explore how these analyses scale with system size in terms of
number of iterations to solution, time to solution, and tokens used. We anticipate
that the simple file storage of analysis data in our current implementation will
become a performance bottleneck. To address that, we plan to integrate the Analysis
Orchestrator with a database to improve journal management and data retrieval. We
also expect that further expansion of \ac{rag} will significantly improve the
performance of \ac{llm} agents.

Our results are indeed very preliminary, but we believe they justify further
investigation in this area.

\section*{Acknowledgments}
This research used resources of the Oak Ridge Leadership Computing Facility at the Oak Ridge National Laboratory, which is supported by the Advanced Scientific Computing Research programs in the Office of Science of the U.S. Department of Energy under Contract No. DE-AC05-00OR22725. Authors thank Eve Tsybina from Oak Ridge National Library for reading an earlier draft and providing feedback that helped us improve the paper.

\bibliographystyle{IEEEtran}
\bibliography{references.bib}

\end{document}